\documentclass{article}
\pdfoutput=1
\PassOptionsToPackage{numbers, compress}{natbib}

\usepackage[final]{neurips_2019_ml4ps}
\usepackage[utf8]{inputenc} 
\usepackage[T1]{fontenc}    
\usepackage{url}            
\usepackage{amsfonts}       
\usepackage{nicefrac}       
\usepackage{microtype}      
\usepackage{array}
\usepackage{comment}
\usepackage{subcaption}

\usepackage[utf8]{inputenc} 
\usepackage[T1]{fontenc}    

\usepackage{url}            

\usepackage{amsfonts}       
\usepackage{nicefrac}       
\usepackage{microtype}      

\usepackage{amsmath,amsthm,amssymb,bm} 
\usepackage{xspace}
\usepackage{array}
\usepackage{enumerate}
\usepackage{enumitem}
\usepackage{stmaryrd}
\usepackage{caption}
\usepackage{subcaption}
\usepackage{multirow}
\usepackage{times}
\usepackage[dvipdfmx]{graphicx}
\usepackage{graphicx} 
\usepackage{algorithm}
\usepackage{hyperref}
\usepackage{sidecap}
\usepackage{wrapfig}
\usepackage{todonotes}
\usepackage{colortbl}
\usepackage{arydshln}
\usepackage{braket}
\numberwithin{equation}{section}

\title{Classical Quantum Optimization with \\Neural Network Quantum States}

\author{%
  Joseph Gomes\\
  University of Iowa\\
  Iowa City, IA 52246\\
  \texttt{joe-gomes@uiowa.edu}\\
  \And
  Keri A. McKiernan\\
  Stanford University\\
  Stanford, CA 94305\\
  \texttt{kmckiern@stanford.edu}\\
  \AND
  Peter Eastman\\
  Stanford University\\
  Stanford, CA 94305\\
  \texttt{peastman@stanford.edu}\\
  \And
  Vijay S. Pande\\
  Stanford University\\
  Stanford, CA 94305\\
  \texttt{pande@stanford.edu}\\
}

\begin{document}

\newcommand\pl[1]{\textcolor{red}{[PL: #1]}}
\maketitle

\begin{abstract}
The classical simulation of quantum systems typically requires exponential resources. Recently, the introduction of a machine learning-based wavefunction ansatz has led to the ability to solve the quantum many-body problem in regimes that had previously been intractable for existing exact numerical methods. Here, we demonstrate the utility of the variational representation of quantum states based on artificial neural networks for performing quantum optimization. We show empirically that this methodology achieves high approximation ratio solutions with polynomial classical computing resources for a range of instances of the Maximum Cut (MaxCut) problem whose solutions have been encoded into the ground state of quantum many-body systems up to and including 256 qubits.

\end{abstract}
\section{Introduction}

The key difficulty in classical quantum simulation is the exponential growth of the Hilbert space dimension with system size, leading to an intractable number of parameters needed for an exact representation of the wavefunction of modest sized many-body systems.  The exponential growth can be partially avoided through the use of approximate methods such as stochastic sampling approaches (Quantum Monte Carlo) \cite{becca2017quantum} or compact representations of the quantum state (tensor network states) \cite{orus2014practical,vidal2004efficient,verstraete2004renormalization}. The neural-network quantum state (NQS) approach proposed by Carleo and Troyer \cite{carleo2017solving} demonstrates the ability of the Restricted Boltzmann Machine (RBM) to compactly represent high dimensional wavefunctions. 

In contrast to previous tensor network based methods, the NQS approach has been shown to capture longer range entanglement structures leading to highly accurate representations of quantum systems. Since the introduction of NQS, there have been numerous applications in condensed matter physics \cite{carleo2018constructing} and strongly correlated electron systems \cite{nomura2017restricted}, fermionic electronic structure simulation \cite{choo2019fermionic}, simulation of general quantum circuits \cite{jonsson2018neural}, quantum state tomography \cite{torlai2018neural}, as well as advances in our understanding of the entanglement properties of the neural network wavefunction ansatz \cite{deng2017quantum,gao2017efficient,chen2018equivalence,clark2018unifying}. For a comprehensive review on applications of NQS, see Ref.~\cite{melko2019restricted}.

One area of quantum simulation where the NQS approach has yet to be applied is quantum optimization algorithms. Challenging computational problems are frequently tackled by classical heuristic algorithms chosen based on empirical performance evaluation \cite{dunning2018works}. 
Optimization problems can also be solved using quantum heuristic algorithms, which typically involves formulating the problem of interest into a "cost Hamiltonian" \cite{farhi2014quantum, hadfield2018representation}. The goal is then to identify the solution, or state, yielding the lowest cost. Universal gate quantum algorithms for optimization commonly minimize this cost Hamiltonian through alternating applications of problem-specific cost and driver Hamiltonians by a series of quantum logic gates \cite{bapat2018bang, hadfield2019quantum}. Adiabatic quantum optimization algorithms minimize this cost Hamiltonian by reversible, adiabatic quantum state evolution between a Hamiltonian whose ground state is easy to prepare and the cost Hamiltonian \cite{albash2018adiabatic}.  

In this work, we demonstrate the utility of the NQS ansatz to obtain the ground state of some quantum many-body systems relevant to quantum optimization. In contrast to universal gate or adiabatic quantum optimization algorithms, we obtain a variational estimate of the solution through direct optimization and subsequent sampling of the NQS wavefunction given a problem-specific cost Hamiltonian. We show empirically that this approach achieves high accuracy solutions with numerically observed polynomial time and memory complexity for a variety of instances of the MaxCut combinatorial optimization problem whose solutions have been encoded into the ground state of a quantum many-body system. The machine learning approach to many-body quantum states allows for approximate simulations of quantum optimization beyond what can be performed exactly on classical computing resources or on current quantum computing resources.

\vspace{-0.2cm}
\section{\label{sec:methods}Methods}

\subsection{Neural Network Quantum States}

We consider a system made of \textit{L} spin-1/2 degrees of freedom denoted as ${\boldsymbol{\sigma}_\textit{j}}$ = $\pm$ 1, \textit{j} = 1, ..., \textit{L}. The amplitude of the (complex) RBM wavefunction is represented by a sum of exponentials,

\begin{equation}
    \Psi(\boldsymbol{\sigma}) = \sum_{\boldsymbol{h}}{e^{\sum_{j}{a_j\sigma_j}+\sum_{i}{b_i h_i}+\sum_{ij}{h_i W_{ij}\sigma_j}}}
\label{eq:rbm1}
\end{equation}

where the sum runs over all $\boldsymbol{h}$ = ($\textit{h}_1$, $\textit{h}_2$,...,\textit{$h_M$}) with the binary variables \textit{$h_i$} $\in$ \{-1,1\} for \textit{i} = 1,...,\textit{M}. By performing the summation over the hidden variables $\boldsymbol{h}$, Eq.~\ref{eq:rbm1} reduces to $log\Psi(\boldsymbol{\sigma}) = \sum_j{a_j\sigma_j} + \sum_i{log[cosh(b_i + \sum_j{W_{ij}\sigma_j})]}$ up to some additive factor which corresponds to the normalization constant and phase factor of the wavefunction.


\subsection{Maximum Cut}
Given an undirected graph $G=(V,E)$ and non-negative weights $w_{ij}$ = $w_{ji}$ on the edges (i, j) $\in$ E, the MaxCut problem seeks to find a bi-partitioning of this graph, such that the sets of vertices $S$ and $\Bar{S}$ maximizes the weight of edges in the cut ($S$, $\Bar{S}$). 
In the case of MaxCut, the cost Hamiltonian minimized is of the form:

\begin{equation}
    \hat{H} = -\frac{1}{4} \sum_{(i,j) \in E} L_{ij} \sigma_i^z \sigma_j^z,
    \label{eq:optH}
\end{equation}

where $L$ is the graph Laplacian and $\sigma^z$ is the Pauli z operator. Note that this problem has a set of solutions given by $2^N$ N-bit bitstrings. The MaxCut combinatorial optimization problem is known to be NP-hard \cite{karp1972reducibility}. 

\subsection{Classical Quantum Optimization}

The energy of a wavefunction $\Psi(\boldsymbol{\sigma})$ can be estimated as
\begin{equation}
\begin{split}
\langle \hat{H} \rangle \approx \left\langle \sum_{\boldsymbol{\sigma}'} \bra{\boldsymbol{\sigma}}\hat{H}\ket{\boldsymbol{\sigma}'} \frac{\Psi(\boldsymbol{\sigma}')}{\Psi(\boldsymbol{\sigma})} \right\rangle_{\boldsymbol\sigma}
\end{split}
\end{equation}
where $\left\langle {}\cdot{} \right\rangle_{\boldsymbol\sigma}$ denotes a stochastic expectation value taken over a sample of configurations $\{\boldsymbol{\sigma}\}$ drawn from the probability distribution corresponding to the variational wavefunction and the summation runs over all configurations $\boldsymbol{\sigma'}$ where the matrix element $\bra{\boldsymbol{\sigma}} \hat{H} \ket{\boldsymbol{\sigma'}}$ is nonzero. Given a variational ansatz $\Psi(\{\alpha_\textit{k}\})$ parameterized by $\alpha_\textit{k}$ for \textit{k} = 1,...,\textit{$N_{var}$}, and target Hamiltonian $\hat{H}$ (Eqn.~\ref{eq:optH}), we optimize using the stochastic reconfiguration (SR) method introduced by Sorella~\cite{sorella1998green}. The SR optimization method can be seen as an effective imaginary-time evolution in the variational subspace. In this view, the optimization procedure may be interpreted as an imaginary-time evolution from a randomly initialized quantum state to one that minimizes the problem-specific cost Hamiltonian given a variational ansatz.

After optimization, we draw a set of $max(1000, 10*N_{var})$ sample configurations and from this set determine the bitstring that minimizes the cost Hamiltonian. 

%
%
%

\subsection{Datasets}
\label{sec:datasets}
We have created a dataset containing 90 instances of fixed edge density ($\rho$ = 12\%, 25\%, 50\%), random graphs of increasing size (|V| = 8, 16, 32, 64, 128, 256) using \texttt{rudy}, a machine independent graph generator written by G. Rinaldi. 
Reference solutions of random graphs containing less than 256 vertices were generated by \texttt{BiqCrunch} \cite{helmberg2000spectral}, a semidefinite branch-and-bound method for solving binary quadratic problems. Due to size limitation of exact branch-and-bound solvers, the BURER2002 algorithm \cite{burer2002rank}, a heuristic based on non-linear optimization with local search, was used, as implemented in \texttt{MQlib} \cite{dunning2018works}, to estimate solutions of random graphs containing 256 vertices.

\subsection{Implementation Details}

All calculations were performed with the free, open source software package \texttt{Netket} \cite{netket:2019}. The following hyperparameters were used: number of hidden variables (\textit{M} = \textit{L} number of spins), learning rate ($\epsilon$=0.05), standard deviation of random initial NQS weights ($\sigma_{sd}$=0.01), optimization steps ($n_{opt}$=200), Monte Carlo samples ($N_{MC}$=$N_{var}$), SR regularization ($\lambda$=0.1). 
Timing benchmarks of the NQS method were obtained by averaging the wall clock time of 10 consecutive optimization steps across all MaxCut instances of a given edge density and number of vertices, except in the case of graphs containing 256 vertices where two instances of a given edge density were used. Timing benchmarks were performed on a single core of a 2.60 GHz 32-core Intel Xeon E5-2650 processor equipped with 256 GB of memory. We note that the hyperparameters used in this study were not optimized for wall clock time. The code and datasets used in this study are available at \href{http://www.github.com/joegomes/cqo}{http://www.github.com/joegomes/cqo}.
\vspace{-0.2cm}
\section{\label{sec:results}Results}

\begin{figure}[h]
    \centering
    \includegraphics[width=\textwidth]{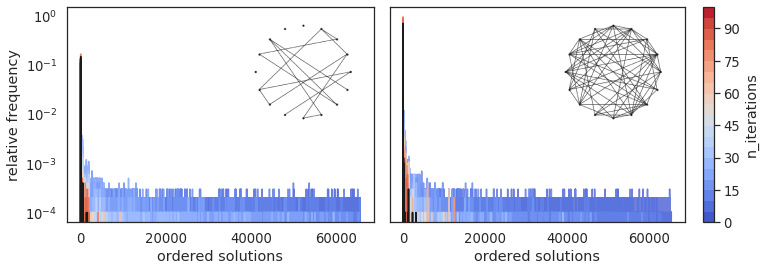}
    \caption{Solving the MaxCut problem via NQS for both sparse (left) and dense (right) 16-node \texttt{rudy} graphs. Each example graph instance is displayed as an inset. The colored distributions illustrate evolution of the reward distribution over the course of the training process. It can be seen NQS optimization procedure gradually increases the probability of sampling the best-quality solutions (indexed 0 and onward), while gradually reducing the probability of sampling the worst-quality solution.
    }
    \label{fig:anneal}
\end{figure}

\begin{figure*}[t!]
    \centering
    \begin{subfigure}[t]{0.45\textwidth}
        \centering
        \includegraphics[height=2.45in]{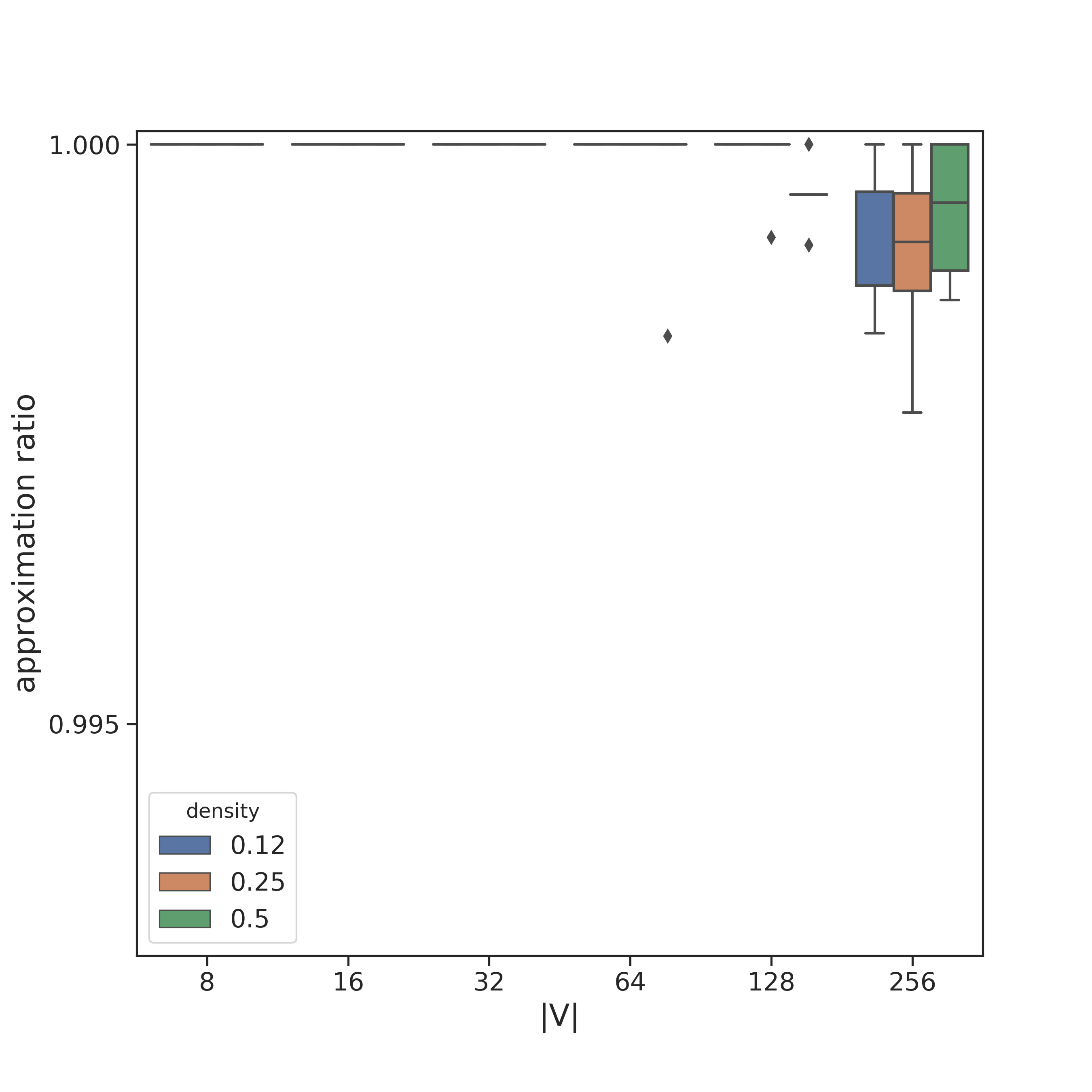}
        \caption{Approximation ratio.}
        \label{fig:approx}
    \end{subfigure}%
    ~ 
    \begin{subfigure}[t]{0.45\textwidth}
        \centering
        \includegraphics[height=2.45in]{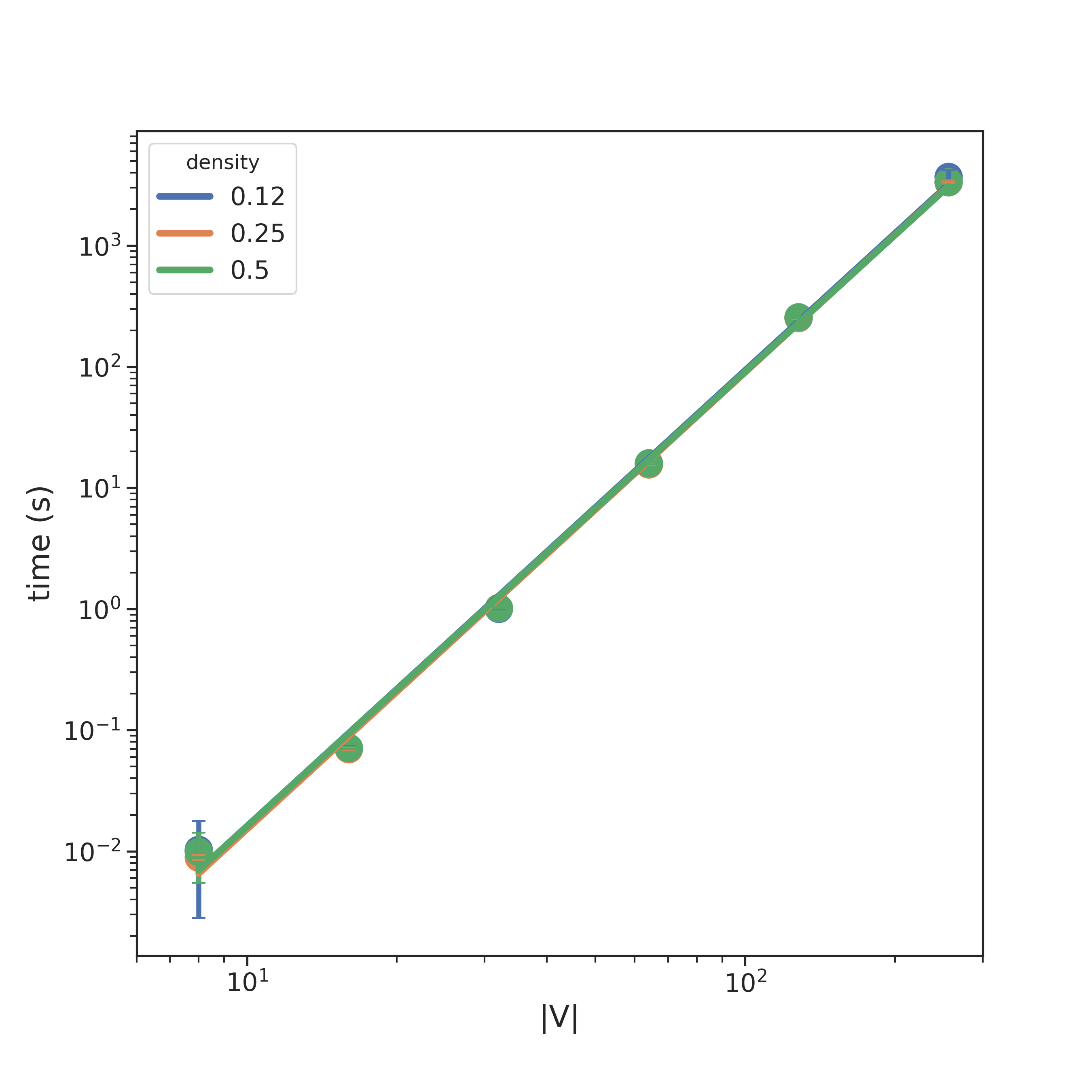}
        \caption{Average time (s) per training step.}
        \label{fig:timing}
    \end{subfigure}
    \caption{Approximation ratio and timing benchmarks for the NQS method on \texttt{rudy} dataset.}
    \label{fig:fig2}
\end{figure*}

Here we apply NQS method to a dataset of MaxCut instances described in Sec.~\ref{sec:datasets}. We first examine performance, with respect to an optimal ordering of solutions, through the progression of the training process (Fig.~\ref{fig:anneal}). Then, we examine quality and time benchmarks across the full \texttt{rudy} dataset (Fig.~\ref{fig:fig2}). For all instances studied here, ground state configurations of NQS wavefunctions were found to generate optimal or nearly optimal rewards. 

Fig.~\ref{fig:anneal} illustrates application of the NQS method to solve the MaxCut problem on two 16-node instances. Following optimization, for a given number of training iterations (specified by the distribution color), 10,000 samples were drawn from the probability distribution corresponding to the NQS wavefunction. These samples were then used to compute an empirical probability distribution over all possible solutions. The optimal ordering of solutions was determined through exhaustive calculation of solution rewards. The final distribution of the optimized NQS wavefunction is displayed in black.

It can be seen that at the beginning of training, the NQS wavefunction yields a nearly uniform reward distribution  (iteration < 20, dark blue distributions). As training progresses, the probability of sampling high-quality solutions increases (red distributions).  For both graph instances, the optimal solution is indeed the most likely solution (index 0, largest peak of the black distribution). In both cases, there remains residual probability of non-optimal solutions. However, note that although not the optimal solution, these solutions are relatively high quality solutions.


Determining the optimality of a solution to a given MaxCut problem requires exponential resources, but an (nearly) optimal solution can be obtained with classical heuristic algorithms \cite{dunning2018works}. In Fig.~\ref{fig:approx}, we compare the solutions obtained by NQS with those from the classical heuristic algorithms described in Sec.~\ref{sec:datasets} by means of the approximation ratio, the multiplicative factor relating the returned and estimated optimal reward. We observed satisfactory performance of the NQS method with an approximation ratio $>$ 0.995 across the set of problems studied without a strong dependence on edge density, although we do not expect this trend to hold in general. We observed that the classical heuristic algorithm BURER2002 outperforms classical quantum optimization in approximation ratio, time to solution, and efficiency. 

Based on the complexity analysis presented in Ref. \cite{carleo2017solving}, we expect a time complexity between $\mathcal{O}(N_{var} * N_{MC})$ and $\mathcal{O}(N_{var}^2 * N_{MC})$ (or $\mathcal{O}(N^4)$ and $\mathcal{O}(N^6)$) for quadratic cost Hamiltonian optimization and observe approximately $\mathcal{O}(N^{4})$ (Fig. \ref{fig:timing}) scaling over the size of systems studied here without a noticeable dependence on graph edge density. The memory complexity for NQS Hamiltonian storage and wavefunction representation ($N_{var}$) grows quadratically with system size when the number of hidden RBM units is related linearly to the number of visible RBM units. 


\vspace{-0.2cm}
\section{\label{sec:conclusion}Outlook}
We have demonstrated the application of the neural-network quantum state approach toward solving quantum optimization problems. We show for the set of instances studied here that this approach achieves accurate classical quantum simulation using polynomial time and memory resources for a variety of MaxCut combinatorial optimization problems, on systems up to 256 qubits in size. 
The construction of a formal proof determining theoretical complexity and performance guarantee remains an open question.
\section*{Acknowledgements}
The Pande Group acknowledges the generous support of Dr. Anders G. Frøseth and Mr. Christian Sundt for our work on machine learning. The Pande Group is broadly supported by grants from the NIH (R01 GM062868 and U19 AI109662) as well as gift funds and contributions from Folding@home donors. J.G acknowledges start up funding from the University of Iowa. This research was supported in part through computational resources provided by The University of Iowa.

V.S.P is a consultant \& SAB member of Schrodinger, LLC and Globavir, sits on the Board of Directors
of Apeel Inc, Asimov Inc, BioAge Labs, Ciitizen, Devoted Health, Freenome Inc, Omada Health,
Patient Ping, Rigetti Computing, and is a General Partner at Andreessen Horowitz.
\bibliography{main}
\bibliographystyle{unsrt}




\end{document}